\documentclass[11pt]{article} 
\usepackage{graphicx}
\usepackage{hyperref}
\usepackage{authblk}
\usepackage{lineno} 
\usepackage{amsmath}
\usepackage{amssymb}
\usepackage{epstopdf}
\usepackage{amsfonts}
\usepackage{color}
\usepackage{bm}
\usepackage{mathrsfs}
\usepackage{url}
\usepackage[normalem]{ulem}
\usepackage{bigints}
\usepackage{afterpage}
\usepackage{array}

\date{\today}

\title{Search for relativistic magnetic monopoles with five years of the ANTARES detector data}

\begin{document} 

\author[a]{A.~Albert}
\author[b]{M.~Andr\'e}
\author[c]{M.~Anghinolfi}
\author[d]{G.~Anton}
\author[e]{M.~Ardid}
\author[f]{J.-J.~Aubert}
\author[g]{T.~Avgitas}
\author[g]{B.~Baret}
\author[h]{J.~Barrios-Mart\'{\i}}
\author[i]{S.~Basa}
% ...


\affil[a]{\scriptsize{GRPHE - Universit\'e de Haute Alsace - Institut universitaire de technologie de Colmar, 34 rue du Grillenbreit BP 50568 - 68008 Colmar, France}}
\affil[b]{\scriptsize{Technical University of Catalonia, Laboratory of Applied Bioacoustics, Rambla Exposici\'o, 08800 Vilanova i la Geltr\'u, Barcelona, Spain}}
\affil[c]{\scriptsize{INFN - Sezione di Genova, Via Dodecaneso 33, 16146 Genova, Italy}}
\affil[d]{\scriptsize{Friedrich-Alexander-Universit\"at Erlangen-N\"urnberg, Erlangen Centre for Astroparticle Physics, Erwin-Rommel-Str. 1, 91058 Erlangen, Germany}}
\affil[e]{\scriptsize{Institut d'Investigaci\'o per a la Gesti\'o Integrada de les Zones Costaneres (IGIC) - Universitat Polit\`ecnica de Val\`encia. C/  Paranimf 1, 46730 Gandia, Spain}}
\affil[f]{\scriptsize{Aix Marseille Univ, CNRS/IN2P3, CPPM, Marseille, France}}
\affil[g]{\scriptsize{APC, Univ Paris Diderot, CNRS/IN2P3, CEA/Irfu, Obs de Paris, Sorbonne Paris Cit\'e, France}}
\affil[h]{\scriptsize{IFIC - Instituto de F\'isica Corpuscular (CSIC - Universitat de Val\`encia) c/ Catedr\'atico Jos\'e Beltr\'an, 2 E-46980 Paterna, Valencia, Spain}}
\affil[i]{\scriptsize{LAM - Laboratoire d'Astrophysique de Marseille, P\^ole de l'\'Etoile Site de Ch\^ateau-Gombert, rue Fr\'ed\'eric Joliot-Curie 38,  13388 Marseille Cedex 13, France}}
% ...

\maketitle 

\begin{abstract}
A search for magnetic monopoles using five years of data recorded with the ANTARES neutrino telescope from January 2008 to December 2012 with a total live time of 1121 days is presented.
The analysis is carried out in the range $\beta$ $>$ $0.6$ of magnetic monopole velocities using a strategy based on run-by-run Monte Carlo simulations.
No signal above the background expectation from atmospheric muons and atmospheric neutrinos is observed, and upper limits are set on the magnetic monopole flux ranging from $5.7 \times 10^{-16}$ to $1.5 \times 10^{-18}$ cm$^{-2} \cdot $ s$^{-1} \cdot $ sr$^{-1}$.\\

\end{abstract}

\section{Introduction\label{sec:1}}

The concept of a particle with a magnetic charge (the magnetic monopole, MM in the following) was introduced by P.~A.~M. Dirac in 1931 \cite{Dirac}  to explain the quantization of the elementary electric charge, $e$. The Dirac basic relation between $e$ and the magnetic charge $g$ is
\begin{equation}\tag{1}\label{gd}
{eg\over c}={n\hbar \over 2 } \quad \longrightarrow \quad g = k \cdot g_D = k\cdot {e\over 2\alpha} \ ,
\end{equation}
where $g_D$ is the unit Dirac charge, $k$ is an integer and $\alpha\simeq 1/137$ is the fine structure constant.
The existence of magnetic charges and currents would symmetrize the Maxwell's equations. However, the symmetry would not be perfect, as $g_D$ is numerically much larger than $e$. 
In 1974, G.~\textquoteright t~Hooft~\cite{hooft} and A.~M.~Polyakov~\cite{polyakov} showed that the electric charge is naturally quantized in Grand Unification Theories (GUTs). MMs appear at the phase transition corresponding to the spontaneous breaking of the unified group into subgroups, one of which is U(1), describing electromagnetism. 

While there is no indication of the mass of the Dirac's magnetic monopole, in the context of GUTs the MM mass $M$ is related to the mass of the $X$-boson carrier of the unified interaction ($m_X \sim 10^{15}$ GeV/c$^2$), yielding $M \gtrsim m_X /\alpha \simeq 10^{17}$ GeV/c$^2$.
MMs with masses $M\sim 10^{5} \div 10^{12}$ GeV/c$^2$ (called intermediate-mass MMs) are predicted by theories with an intermediate energy scale between the GUT and the electroweak scales and would appear in the early Universe at a considerably later time than the GUT epoch \cite{laza-imm}.
More recently, it has been proposed \cite{cho} that solutions yielding MMs could arise within the electroweak theory itself. This Cho-Maison, or {electroweak MM}, would be expected to have a mass of the order of several TeV.

Guided mainly by Dirac's argument and their predicted existence from spontaneous symmetry breaking mechanisms, searches have been routinely made for MMs produced at accelerators, in cosmic rays, and bound in matter \cite{rpp,maurizio}. Eq. (\ref{gd}) defines most of the MM properties, as they are assumed as point-like particles, of magnetic charge equal $g$, with unknown mass and with unknown relic cosmic abundance. To date, there are no confirmed observations of exotic particles possessing magnetic charge.

MMs at the electroweak scale with $M<10$ TeV are very good candidates for searches at the CERN Large Hadron Collider (LHC). The ATLAS collaboration~\cite{aad} searched for MMs as highly ionizing particles produced in proton-proton collisions, leading to new cross section upper limits for spin 1/2 and spin 0 particles. MoEDAL is a dedicated experiment searching for MMs produced in high-energy collisions at the LHC using stacks of nuclear-track detectors and a trapping detector. Recently, limits on MM production cross sections have been reported both for the 8~TeV and 13~TeV LHC runs~\cite{acharya,acharya2}.

GUT MMs are very massive and composite objects, well beyond the reach of any existing or foreseen accelerator. They could have been produced in a phase transition in the early Universe \cite{kibble}, and appeared as topological defects, about one pole for each causal domain. 
This would lead to a present-day overabundance \cite{pre79}: the reduction of the number of MMs in the Universe was one of the motivating factors for cosmological inflation in Guth's original work \cite{guth}.

As the Universe expanded and cooled down, the energy of MMs decreased: they would have reached a speed $\beta=v/c \sim 10^{-10}$ during the epoch of galaxy formation ($v$ is the MM speed and $c$ is the speed of light in vacuum).
After the gravitationally-driven galaxy formation epoch, galactic magnetic fields developed through the dynamo mechanism. Then, MMs were re-accelerated by these magnetic fields, yielding an isotropic intergalactic flux of relatively high-energy MMs.
A magnetic field $\mathbf{B}$ acting over a length $\ell$ increases the MM kinetic energy by a quantity $gB \ell$. 
The final speed depends on galactic magnetic field strength, on the coherent length $\ell$ and on MM mass and magnetic charge. For the typical values in our Galaxy, i.e. $B \sim 3\ 10^{-6}$ G and $\ell \sim 300 \textrm{ pc} = 10^{21}$ cm, MMs of $g=g_D$ are relativistic up to $M\sim 10^{11}$ GeV/c$^2$. Then, their velocity decreases to reach the value $\beta\simeq 10^{-3}$ for $M\gtrsim 10^{17}$ GeV/c$^2$.
 In models in which the cosmic magnetic field, instead of being uniformly distributed, is strongly correlated with the large scale structure of the universe, MMs are relativistic up to $\sim 2 \times 10^{13}$~GeV/c$^2$ for $g=g_D$~\cite{ryu}.

The above MM acceleration process drains energy from the galactic magnetic field. An upper bound on the flux of MMs in the galaxy (called the Parker bound \cite{parker}) has been obtained by requiring the rate of this energy loss to be small compared to the time scale on which the galactic field can be regenerated. With reasonable choices for the astrophysical parameters \cite{rpp}, the Parker bound corresponds to
\begin{equation}\tag{2}\label{eq:2.din6}
\Phi_M  \lesssim \left\{
\begin{array}{ll}
10^{-15} \quad\quad\quad\quad\quad \textrm{ [cm}^{-2} \textrm{s}^{-1} \textrm{sr}^{-1}],\quad & M \lesssim 10^{17} \textrm{ GeV(c}^2 \\
10^{-15} \big( { 10^{17}\textrm{ GeV}\over M }\big) \ \ \textrm{ [cm}^{-2} \textrm{s}^{-1} \textrm{sr}^{-1}],\quad  & M \gtrsim 10^{17} \textrm{ GeV/c}^2
\end{array}\right.
\end{equation}

Search strategies are determined by the expected interactions of MMs as they pass through matter. These would give rise to a number of peculiar signatures. A complete description of the techniques used for the search of these particles is in \cite{maurizio}, and a complete list of the results in \cite{rpp}.

Several searches were carried out also using neutrino telescopes. The ANTARES neutrino telescope~\cite{ageron} was completed in 2008 and the collected data can be used to search for MMs with energies high enough to yield light emission. The results of the analysis published in~\cite{adrian} using a data set of 116 days live time, lead to upper limits on the flux in the range between $1.3 \times 10^{-17}$ and $5.7 \times 10^{-16}$ cm$^{-2} \cdot$ s$^{-1} \cdot$ sr$^{-1}$ for MMs with $\beta >0.6$.
The IceCube collaboration has set upper limits on the flux for relativistic MMs ranging from $1.55 \times 10^{-18}$ to $10.39 \times 10^{-18}$ cm$^{-2} \cdot$ s$^{-1} \cdot$ sr$^{-1}$~\cite{anna}.

In this paper, a new analysis is presented, based on an enlarged ANTARES data set of 1121 days collected from 2008 to 2012, increasing by a factor of $\sim$10 the live time of the previous published result. This analysis is based on a new selection of cuts, yielding a better separation of the MM signal from the background of atmospheric muons and neutrinos. Further, it relies on a new simulation strategy that reproduces each data run individually, allowing for an accurate reproduction of the data taking conditions. 

The paper is organized as follows: a brief description of the ANTARES telescope and the MM expected signatures are given in sections~2 and~3, respectively. The simulation and reconstruction algorithms are described in sections~4 and~5. The MM-sensitive observables, the selection strategy and the upper limit calculation are discussed in sections~6 and~7. Finally, the results are presented and discussed in section~8.

\section{The ANTARES telescope}
The ANTARES detector~\cite{ageron} is an undersea neutrino telescope anchored 2475~m below the surface of the Mediterranean Sea and 40~km offshore from Toulon (France). It consists of 12 detection lines with 25 storeys per line and 3 optical modules (OMs) with 10-inch photomultipliers (PMTs) per storey. The detection lines are 450~m long and spaced 60$-$75~m apart horizontally. The main channel for neutrino detection is via the muons produced from high-energy muon neutrinos interacting inside, or in the vicinity of the detector. These muons move at relativistic velocities and induce the emission of Cherenkov light along their paths, detected by the optical modules.
PMT signals corresponding to a charge above a threshold of 0.3~photo-electrons are integrated with a time window of 40~ns, digitised and denoted as hits. The readout of OMs is performed in the storey\textquoteright s Local Control Module, which collects the data in packages of 104 ms. These packages are sent to an on-shore farm of computers for further data processing and filtering.
Each detector storey has one local clock that is synchronized to the on-shore master clock~\cite{daq}. Furthermore, at the computer farm a system of triggers is applied on the data (see section~5), selecting signatures which may correspond to the passage of relativistic particles.

\section{Detection of magnetic monopoles}
The signature of a MM in a neutrino telescope like ANTARES is similar to that of a highly energetic muon. Thus, as in the case of electrically-charged particles, magnetically-charged particles induce the polarization of the dielectric medium. Coherent light emission (Cherenkov effect) is induced by the restoring medium if the particle travels with a speed above the Cherenkov threshold $\beta_{th}=1/n$, where $n$ is the refractive index of the medium~\cite{tompkins}. In water the threshold is $\beta_{th} \approx 0.74$. The number of photons emitted from a MM with magnetic charge $g$ in a small interval of path length, $dx$, and in the range $d\lambda$ of wavelength, for $\beta \geq \beta_{th}$ can be expressed as
\begin{equation}\tag{3}\label{F1}
\frac{d^{2}n_\gamma}{d\lambda dx} =\frac{2 \pi \alpha}{\lambda^{2}}\left(\frac{n g}{e}\right)^{2} \left(1-\frac{1}{n^{2} \beta^{2}}\right),
\end{equation}
where $n_\gamma$ is the number of photons emitted and $\lambda$ is their wavelength; the remaining quantities are already defined in Eq.~(\ref{gd}). For a given velocity, the Cherenkov radiation yield by a MM is a factor $\left(\frac{n g}{Z e}\right)^{2}$ larger than that from a particle with electric charge $Z e$. 
Thus, for the refractive index of sea water, fast MMs with $g=g_D$ are expected to emit about 8550 times more Cherenkov photons than relativistic muons.

\begin{figure}[hbtp]
\centering
\includegraphics[scale=0.5]{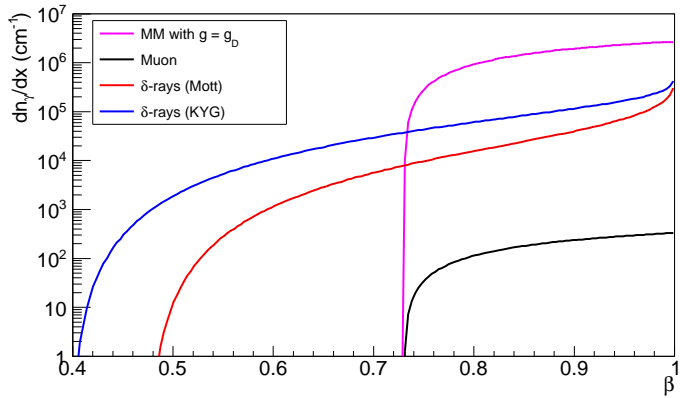} 
\caption{The total number of Cherenkov photons with wavelengths between 300 and 600 nm that are directly produced per centimeter path length by a MM with $g=g_D$, as a function of its velocity ($\beta$). The number of photons produced by $\delta$-rays with Mott cross section model~\cite{ahlen1} and KYG cross section model~\cite{kyg} and by a minimum ionizing muon are also shown.
\label{light}}
\end{figure}

In addition to the direct Cherenkov radiation, MMs can knock off atomic electrons ($\delta$-ray electrons) that can have velocities above the Cherenkov threshold, contributing to the total light yield.
The production of $\delta$-electrons is described by the differential cross-section of Kasama, Yang and Goldhaber (KYG)~\cite{kyg} or by the more conservative (in terms of photon yield) Mott cross section~\cite{ahlen1}.
The contributions to the light yield from these mechanisms are shown in Fig.~\ref{light}. In both cases, some commonly accepted assumptions for the quantum-mechanical aspects of the interaction between a MM and an electron are used that must be implemented in the simulations. 
In this work, the Mott cross section is used, starting for the minimum velocity of $\beta=0.5945$: this allows a simpler application in the Monte Carlo simulation of the spectrum of the produced $\delta$-ray electrons, yielding a safer estimate of the light yield.
Contributions from radio-luminescence of water, pair production, Bremsstrahlung and photo-nuclear reactions induced by relativistic MMs are negligible compared to the direct and indirect Cherenkov light presented in Fig.~\ref{light}, and are not taken into account in this analysis.

In neutrino telescopes, the background of atmospheric muons dominates the solid angle region corresponding to down-going events. In particular, muons in bundle can easily be misidentified with the passage of a relativistic highly ionizing particle. On the opposite, the solid angle region corresponding to up-going events is almost background free, apart from the events induced by atmospheric neutrinos and the surviving down-going atmospheric muons misreconstructed as up-going. Due to the energy spectrum of atmospheric muon neutrinos, they usually induce minimum ionizing muons that can be easily distinguished from fast MMs. In order to suppress the irreducible background of atmospheric muons, only up-going MMs were considered.

The request of up-going MMs reduces the range of masses $M$ that can be observed in a neutrino telescope. The stopping-power defined by S.~P.~Ahlen~\cite{ahlen2} has been used to estimate the absorption and energy loss of a MM when crossing the Earth. This work
has established for MMs the equivalent of the Bethe-Bloch formula that describes the energy loss in the passage of a heavy electric charge by ionization and excitation in a non-conductive medium. Thus, the stopping-power of a MM crossing the Earth could be estimated using the simplified density profile established by Derkaoui et al.~\cite{derkaoui}. Despite the high energy loss, MMs would remain relativistic and detectable as up-going events if $M\gtrsim 10^{10}$~GeV/$c^{2}$ (see for instance Fig.~3 of~\cite{maurizio}). 
As discussed in Section \ref{sec:1}, the MM speed depends on the characteristic of the galactic magnetic fields and on the mass $M$. 
Within reasonable astrophysical considerations, only MMs with a mass $M \lesssim 10^{14}$~GeV/$c^{2}$ can be expected in neutrino telescopes as an up-going event with a speed exceeding the Cherenkov threshold. Thus, the limits presented in this paper hold for MM in the mass range $10^{10}$~GeV/$c^{2} \lesssim M  \lesssim 10^{14}$~GeV/$c^{2}$.

\section{Monte Carlo simulation}
In this section, the simulation of the MM signal and the atmospheric (neutrino and muon) background events are discussed.

\subsection{Magnetic monopole simulation}
Up-going MMs with one unit of Dirac charge, $g=g_D$, have been simulated using nine equal width ranges of velocity in the region $\beta=[0.5945,0.9950]$. The nine intervals of the velocity are defined in the first column of Table~1.

MMs have been simulated using a Monte Carlo program based on GEANT3~\cite{soft}. The simulation is independent of the MM mass and the incoming direction of MMs was distributed isotropically over the lower hemisphere.
The propagation and detection of emitted photons is processed inside a virtual cylindrical surface surrounding the instrumented volume around the detector. A radius of 480 m is chosen to take into account the large amount of light emitted by MMs. 

\subsection{Background simulation}
The main source of background comes from up-going muons induced by atmospheric neutrinos and down-going atmospheric muons wrongly reconstructed as up-going tracks. The simulation of atmospheric muons is carried out using the generator MUPAGE~\cite{carminati} based on the parametrisation of the angle and energy distributions of muons under-water as a function of the muon bundle multiplicity~\cite{becherini}. MUPAGE produces muon events on the surface of the virtual cylinder.

Up-going atmospheric neutrinos from the decay of pions and kaons are simulated using the package GENHEN~\cite{mar,marg} assuming the model from the Bartol group~\cite{aqrawal,barr} which does not include the decay of charmed particles.
The analysis presented in this paper is based on a run-by-run Monte Carlo simulation~\cite{rbr}, which takes into consideration the real data taking conditions of the detector (e.g. sea water conditions, bioluminescence variability, detector status).

\section{Trigger and reconstruction}
The applied triggers are based on local coincidences defined as the occurrence of either two hits on two separate optical modules of a single storey within 20~ns, or one single hit of large amplitude, typically more than 3 photo-electrons.
The trigger used for this analysis is defined as a combination of two local coincidences in adjacent or next-to-adjacent storeys within 100~ns or 200~ns, respectively. In this analysis, only events passing such a trigger, well suited for MMs, are considered.

The event reconstruction has been done with a slightly modified version of the algorithm described in~\cite{aguilar}. By default, it assumes that particles travel at the speed of light. In order to improve the sensitivity for MMs travelling with lower velocities, the algorithm was modified such as to leave the reconstructed velocity of the particle $\beta_{fit}$ as a free parameter to be derived by the track fit.

The algorithm performs two independent fits: a track fit and a bright-point fit. The former reconstructs particles crossing the detector, while the latter reconstructs showering events, as those induced by $\nu_e$ charged current interactions.
Both fits minimize the same $\chi^2$ quality function, thus, two parameters defining the quality of these reconstructions are introduced, $t\chi^2$ for the track fit, and $b\chi^2$ for the bright-point fit.

Some basic quality cuts have been applied to the data to ensure good data taking conditions~\cite{sparking}. To avoid any experimental bias, the search strategy is based on a blind analysis. The selection cuts applied on the analysis are established on Monte Carlo simulations and using a test data sample of about 10\% of the total data set, equivalent to 109 days out of the total 1121 days of live time. These runs are not used later for setting the limits. 

\begin{figure}[hbtp]
\centering
\includegraphics[scale=0.6]{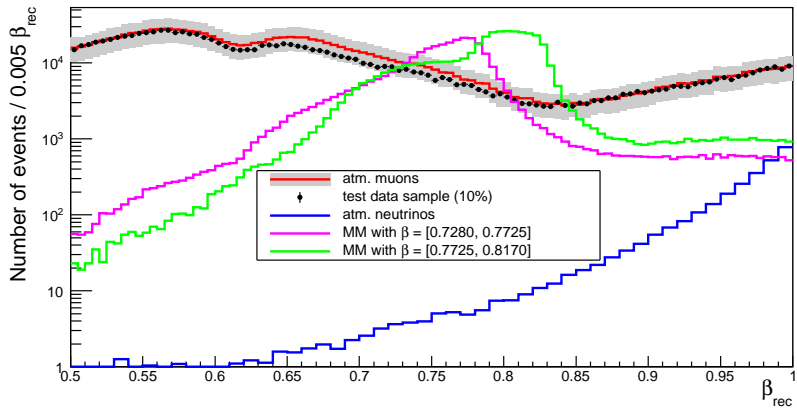} 
\caption{The distribution of the reconstructed $\beta_{fit}$ for atmospheric muons (red histogram) with an uncertainty band of 35\% (filled in gray), atmospheric neutrinos (blue histogram)  and data (points with error bars). For comparison, the distributions of the reconstructed $\beta_{fit}$ for MMs simulated in the velocity ranges $[0.7280, 0.7725]$ (magenta histogram) and $[0.7725, 0.8170]$ (green histogram) are also shown. All distributions correspond to events reconstructed as up-going.
\label{beta}}
\end{figure}

In the following comparisons between the test data sample and simulation, the full collection of Monte Carlo runs is used, and the 10\% of test data is scaled to the total live time.
Fig.~\ref{beta} shows the distribution of the reconstructed velocity $\beta_{fit}$ for MM events, atmospheric muons and neutrinos and compared to the test data sample. The neutrino distribution represents electron neutrinos and muon neutrinos for both neutral and charged currents.

\section{Event selection}

\begin{figure}[hbtp]
\centering
\includegraphics[scale=0.6]{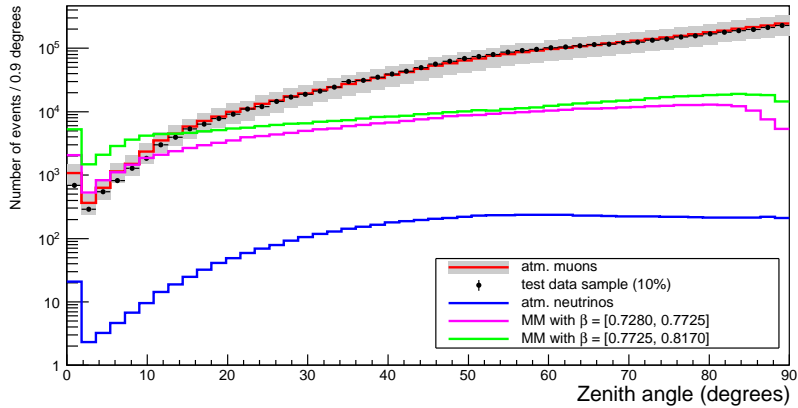} 
\caption{Reconstructed zenith angle for atmospheric muons (red histogram) with an uncertainty band of 35\% (filled in gray), atmospheric neutrinos (blue histogram) and data (points with error bars). For comparison, the distributions of the reconstructed zenith angle for MMs simulated in the velocity ranges $[0.7280, 0.7725]$ (magenta histogram) and $[0.7725, 0.8170]$ (green histograms) are also shown. The peak at zenith = 0$^{\circ}$ represents wrongly reconstructed events.
\label{zenith}}
\end{figure}

In order to remove the bulk of down-going events, only up-going events with reconstructed zenith angles $\leq$ 90$^\circ$ are selected (Fig.~\ref{zenith}). Thus, the comparison shows a good agreement between the test data sample and simulation.
The systematic uncertainties affecting the predictions of atmospheric neutrino and atmospheric muon fluxes are discussed in section~8. Accordingly, the event distributions of these two channels shown in this paper suffer from an overall normalization uncertainty of about 30\% and 35\%, respectively.

Additional cuts on the track fit quality parameter are implemented to remove misreconstructed atmospheric muon tracks. In particular, the requirement $t\chi^2 \leq b\chi^2$ is applied to favour events reconstructed as a track rather than those reconstructed as a bright point.
The further event selections were optimized for different MM velocities. A different event selection was performed for each of the nine bins of $\beta$ reported in the first column of Table~1.

The modified reconstruction algorithm which treats $\beta_{fit}$ as a free parameter was used only in the regions of low velocities between $\beta=0.5945$ and $\beta=0.8170$ (five bins). Thus, MMs with these velocities could be distinguished from particles traveling with the speed of light ($\beta_{fit}=1$). For each of the five low beta bins, only events reconstructed with $\beta_{fit}$ in the range of simulated $\beta$ were used in the final selection. For example, at the range $\beta=[0.5945, 0.6390]$, only events with reconstructed velocity $\beta_{fit}=[0.5945, 0.6390]$ were selected.
In the high velocity interval ranging from $\beta=0.8170$ to $\beta=0.9950$ (four bins), the $\beta_{fit}$ is not a discriminant variable anymore. However, MMs emit a large amount of light compared to that emitted from other particles, which allows them to be distinguished.

\begin{figure}[hbtp]
\centering
\includegraphics[scale=0.6]{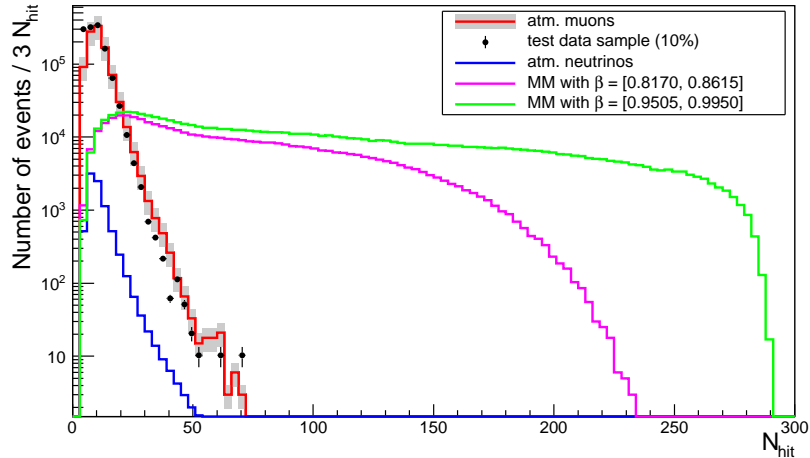} 
\caption{$N_{hit}$ distribution for atmospheric muons (red histogram) with an uncertainty band of 35\% (filled in gray), atmospheric neutrinos (blue histogram) and data (points with error bars). For comparison, the distributions of $N_{hit}$ for MMs simulated in the velocity ranges $[0.8170, 0.8615]$ (magenta histogram) and $[0.9505, 0.9950]$ (green histogram) are also shown. At high velocities,  $N_{hit}$ provides a good discrimination for MM signals after applying the cuts zenith $\leq 90^\circ$ and $t\chi^2 \leq b\chi^2$.
\label{nhit}}
\end{figure}

In the used reconstruction algorithm, the hits from the optical modules belonging to the same storey are summed together to form a \textit{track hit}. The coordinates of its position are coincident with the center of the storey, the time is equal to the time of the first hit and the charge equal to the sum of the hits charges.
For all velocity bins, the number of storeys with selected track hits $N_{hit}$, is used as a powerful discriminant variable since it refers to the amount of light emitted in the event (see Fig.~\ref{nhit}).

\begin{figure}[hbtp]
\centering
\includegraphics[scale=0.6]{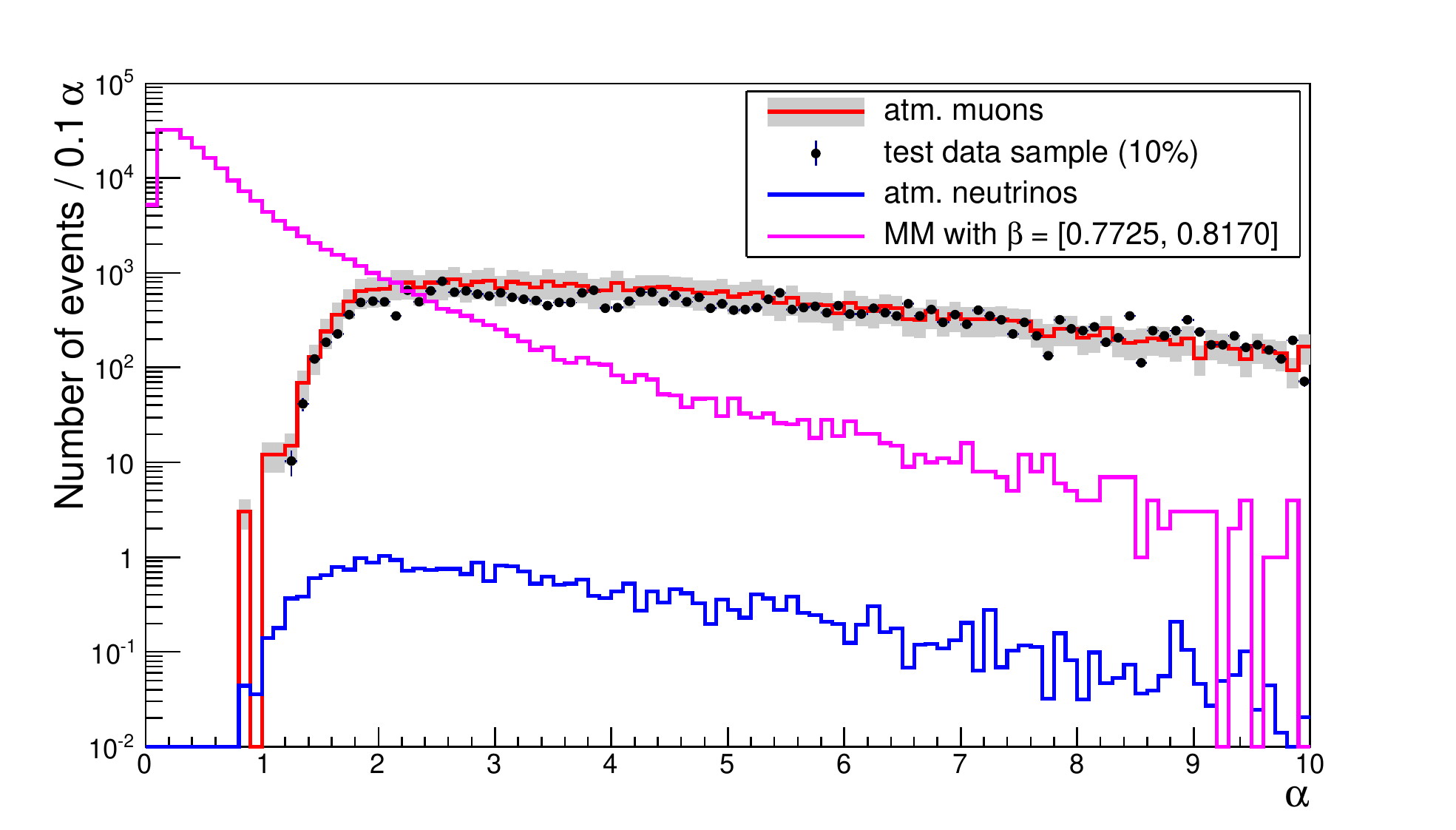} 
\caption{Distribution of the $\alpha$ variable for atmospheric muons (red histogram) with an uncertainty band of 35\% (filled in gray), atmospheric neutrinos (blue histogram) and data (points with error bars). For comparison, the distribution of the $\alpha$ variable for MMs simulated in the velocity range $[0.7725, 0.8170]$ (magenta histogram) is also shown. Only events with reconstructed velocity $\beta_{fit}=[0.7725, 0.8170]$ were selected, and the cuts zenith $\leq 90^\circ$ and $t\chi^2 \leq b\chi^2$ have been applied.
\label{alpha}}
\end{figure}

A second discriminative variable is introduced to further reduce the background, in particular for the velocities below the threshold for direct Cherenkov radiation where the light emission is lower. This variable, named $\alpha$, is defined from a combination of the track fit quality parameter $t\chi^2$ and $N_{hit}$ following~\cite{aguilar}:

\begin{equation}\tag{4}\label{F5}
\alpha=\frac{t\chi^2}{1.3+\left(0.04\times(N_{hit}-N_{df})\right)^2}  ,
\end{equation}
where $N_{df}$ is the number of free parameters in the reconstruction algorithm. It is equal to 6 when $\beta_{fit}$ is included in the reconstruction, and 5 when the velocity is not reconstructed. Example of $\alpha$ distribution is shown at Fig.~\ref{alpha}. This parameter has the advantage of including the track fit quality parameter balanced with the brightness of the events, avoiding that bright events get cut by the condition applied on the $t\chi^2$ variable.

\section{Optimization of cuts}
\begin{figure}[hbtp]
\hspace*{-1.5cm}
\centering
\includegraphics[scale=0.33]{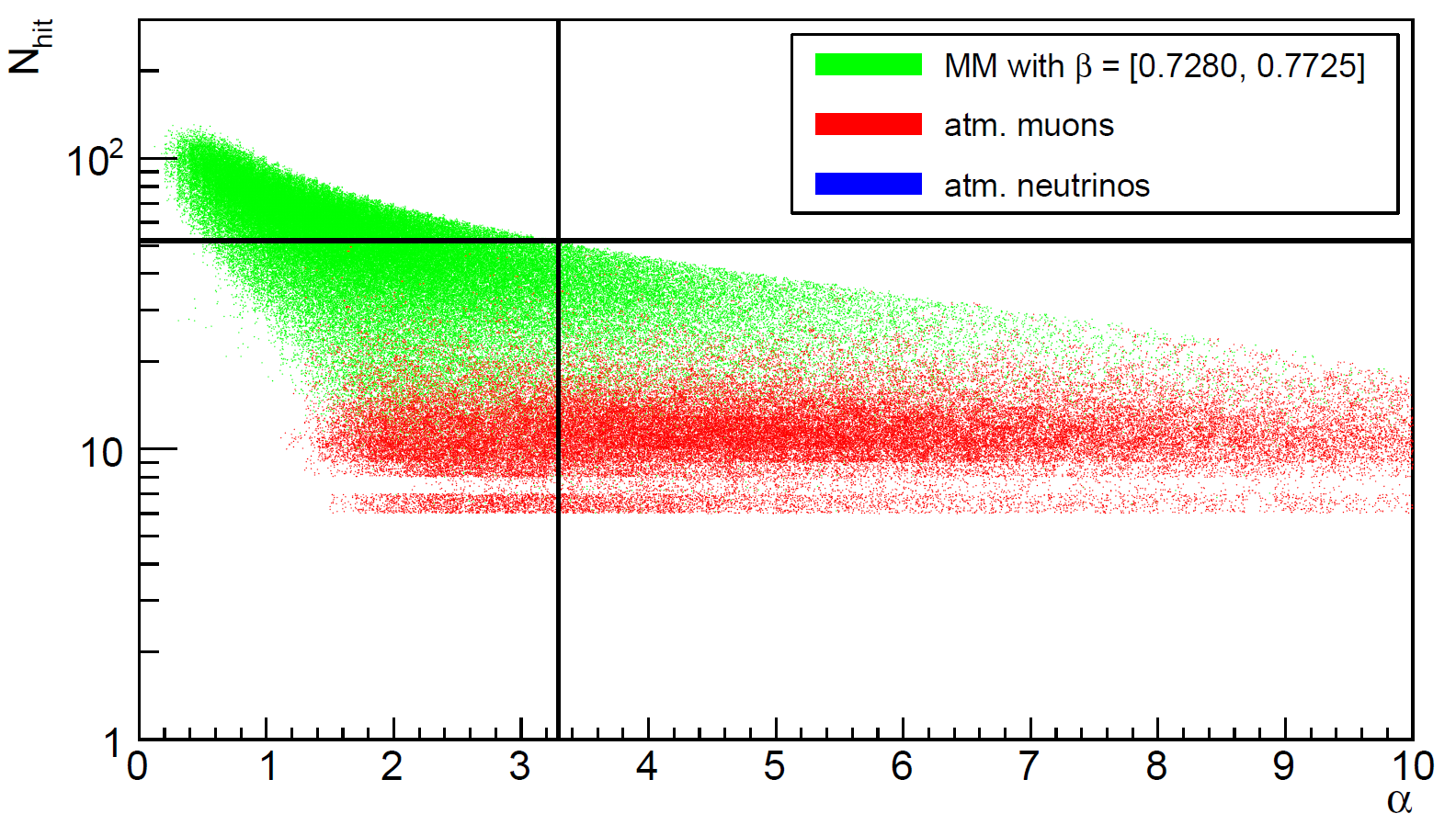}
\caption{Two-dimensional distribution of $\alpha$ and $N_{hit}$, for atmospheric muons, atmospheric neutrinos, and MMs simulated in the velocity range $[0.7280, 0.7725]$. The cuts zenith~$\leq 90^\circ $ and $t\chi^2 \leq b\chi^2$ have been applied, as well as the cut $\beta_{fit} = [0.7280, 0.7725]$. The vertical and horizontal lines show the cuts applied after optimization. No neutrinos survived at this range of $\beta$.
\label{anhit}}
\end{figure}

The following step to suppress the atmospheric background is to use specific cuts on the $N_{hit}$ and $\alpha$ parameters in order to maximize the signal-to-noise ratio. In Fig.~\ref{anhit}, the event distribution of $\alpha$ as a function of $N_{hit}$ is shown for one range of MM velocity. This distribution indicates that a good separation of MM signal from background is achievable. The horizontal and vertical lines show the effect of the cuts. The signal region corresponds to the left upper quadrant.

The 90\% confidence level interval $\mu_{90}(n_b,n_{obs})$, where $n_b$ is the number of background events is the 90\% confidence interval defined by the Feldman-Cousins approach~\cite{feldman}. It depends on the number of observed events $n_{obs}$ which is not known at this point because of the blind approach. Instead, the average confidence interval $\bar{\mu}_{90}(n_b)$ is calculated, from which the sensitivity of the analysis can be derived, by assuming a Poissonian probability distribution for the number of observed events $n_{obs}$.
The selection cuts are optimized by minimizing the so-called Model Rejection Factor (MRF)~\cite{picot}: 
\begin{equation}\tag{5}\label{F6}
MRF=\frac{\bar{\mu}_{90}(n_b)}{n_{MM}} ,
\end{equation}
where $n_{MM}$ is the number of signal events remaining after the cuts, assuming an isotropic MM flux with $\phi^0_{MM}=1.7\cdot10^{-13}$~cm$^{-2} \cdot $~s$^{-1} \cdot $~sr$^{-1}$. In addition to the specific values of the cuts, $n_{MM}$ depends on the detector acceptance $S_{eff}$ ($cm^{2} \cdot sr$) and on the time period over which data was collected $T(s)$.

\begin{figure}[hbtp]
\centering
\includegraphics[scale=0.6]{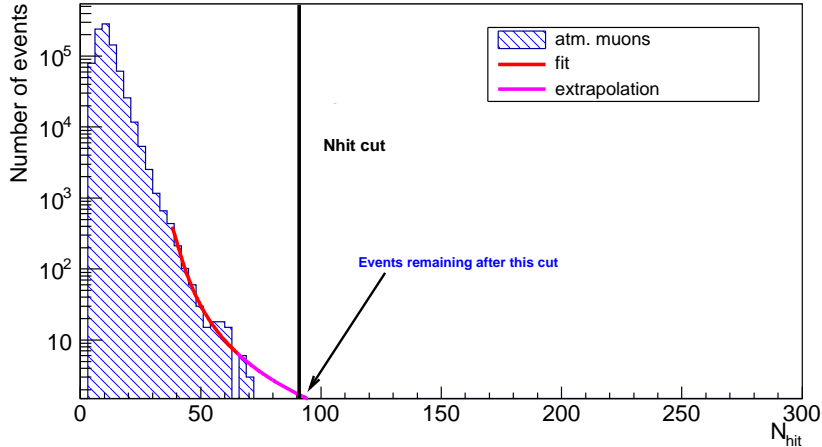} 
\caption{The distribution of $N_{hit}$ for atmospheric muons, extrapolated using a Landau fit function. The contribution of the extrapolation in the total number of events was taken into account in the optimization and the extrapolation uncertainties were computed. For this bin $\beta=[0.8170,0.8615]$, 1.4 events are found after the cut $N_{hit} \geqslant 91$.
\label{extra}}
\end{figure}

In order to compensate for the lack of statistics in the remaining sample of atmospheric muon background, an extrapolation has been performed in the region of interest for the signal. An example of extrapolation performed is shown in Fig.~\ref{extra}. After fitting the $N_{hit}$ distribution for muons with a Landau type function (red), the latter is extrapolated to the region of interest (pink), then the number of muons remaining after the final cut on $N_{hit}$ is given by the sum of the events from the muon histogram (blue) and the extrapolation (pink).
Columns 3 and 4 of Table~1 shows the background expectation, dominated by atmospheric muons, for each bin of $\beta$. After the optimization procedure and the estimation of the background, the 90\% confidence level upper limit on the MM flux is obtained from the values of the cuts yielding the minimum value of the Model Rejection Factor MRF:
\begin{equation}\tag{6}\label{F7}
 \phi_{90\%} = \phi^0_{MM} \cdot MRF.
\end{equation}

\section{Results and discussion}
The unblinding was performed on the total set of data collected by the ANTARES telescope during five years, which corresponds to 1012 active days live time after subtracting the 10\% burn sample. No significant excess of data events is observed over the expected background, and the upper limits on flux have been found using Eq.~(\ref{F7}). Table~1 summarizes, for each of the nine bins of $\beta$, the selection cuts, the number of expected background and observed events, and the 90\% C.L. upper limits on the MM flux. 

The computation of the 90\% C.L upper limits through Eq.~(\ref{F7}) includes the statistical uncertainties on the expected atmospheric muon events in column 3 of Table~1. These uncertainties are dominant over the uncertainties related to the detector response.
The effects on the muon and neutrino rates due to the detector uncertainties are widely discussed elsewhere, particularly in~\cite{sparking,eff,am,sam}. For the atmospheric neutrinos, the systematic uncertainties as a function of the energy are detailed in~\cite{am}. As shown in Table~1, the contribution of atmospheric neutrinos is almost negligible with respect to atmospheric muons and the effects of these uncertainties have been ignored.
Concerning atmospheric muons, the dominant detector effects are connected to the angular acceptance of the optical module~\cite{amram} and to the absorption and scattering lengths in water~\cite{ja}. The maximum $\pm 15\%$ uncertainty on the optical module acceptance and the $\pm 10\%$ on the light absorption length in water over the whole wavelength spectrum yields an overall $^{+35\%}_{-30\%}$ effect on the expected muon rate~\cite{eff}. However, as already stated, in this case the dominant effect (in most cases, with effects larger than $\pm 50\%$ on the number of events) is due to the lack in the statistics of the surviving muons and to the procedure for the background extrapolation, as described in Fig.~\ref{extra}. The values reported in column 3 represent the overall uncertainties on the surviving muon background in each $\beta$ bin.
 
The effect of a third uncertainty, due to the use of the Mott cross-section instead of the KYG (as discussed in Section~3) has not been considered. In this case, a more conservative choice in terms of photon yield has been made. The outcome is to neglect a possible larger photon yield, that has the effect of decreasing the detection thresholds towards smaller values of $\beta$ in Fig.~\ref{light}.
{\renewcommand{\arraystretch}{1.3}
{\setlength{\tabcolsep}{0.1cm}
\begin{table}[htbp]
\begin{center}
\hspace*{-1.5cm}
\begin{tabular}{|c|*{2}{c}|c|c|c|c|}
  \hline
 
\footnotesize $\beta$ range & \multicolumn{2}{c|}{\footnotesize Selection cuts} & \footnotesize Number of &\footnotesize Number of & \footnotesize Number of & \footnotesize Flux Upper Limits \\
& \footnotesize $\alpha$ & \footnotesize $N_{hit}$& \footnotesize atm. muons &\footnotesize atm. neutrinos & \footnotesize obs. events & \footnotesize  90\% C.L. (cm$^{-2} \cdot$ s$^{-1} \cdot$ sr$^{-1}$) \\ 

 \hline  \hline

  \footnotesize [0.5945, 0.6390] &\footnotesize $< 5.5$ &\footnotesize $\geqslant 36$ & \footnotesize 1.9 $\pm$ 0.8 & \footnotesize 1.6 $\times 10^{-4}$ & 0 & \footnotesize $5.9 \times 10^{-16}$ \\

 \hline

\footnotesize [0.6390, 0.6835] &\footnotesize $< 5.0$ &\footnotesize $\geqslant 39$ & \footnotesize 0.9 $\pm$ 0.5 & \footnotesize 1.5 $\times 10^{-4}$&	\footnotesize 0 &	\footnotesize $3.6 \times 10^{-17}$ \\

\hline

\footnotesize [0.6835, 0.7280] &\footnotesize $< 3.4$ &\footnotesize $\geqslant 51$ & \footnotesize 0.9 $\pm$ 1.0 & \footnotesize 1.2 $\times 10^{-4}$&\footnotesize 0 &	\footnotesize $2.1 \times 10^{-17}$ \\

\hline

\footnotesize [0.7280, 0.7725] &\footnotesize $< 3.3$ &\footnotesize $\geqslant 51$ & \footnotesize 1.1 $\pm$ 0.5 & \footnotesize 9.3 $\times 10^{-3}$&\footnotesize 1 &	\footnotesize $9.1 \times 10^{-18}$ \\

\hline

\footnotesize [0.7725, 0.8170] &\footnotesize $< 1.8$ &\footnotesize $\geqslant 73$ & \footnotesize 0.6 $\pm$ 0.4 & \footnotesize 1.0 $\times 10^{-3}$&\footnotesize 0 &\footnotesize $4.5 \times 10^{-18}$ \\

\hline  \hline

\footnotesize [0.8170, 0.8615] &\footnotesize $< 0.8$&\footnotesize $\geqslant 91$ & \footnotesize 1.4 $\pm$ 0.9 & \footnotesize 1.8 $\times 10^{-1}$&\footnotesize 1 &	\footnotesize $4.9 \times 10^{-18}$ \\

\cline{1-5} \cline{7-7}

\footnotesize [0.8615, 0.9060] &\footnotesize $< 0.6$&\footnotesize $\geqslant 92$ & \footnotesize 1.3 $\pm$ 0.8 & \footnotesize 1.6 $\times 10^{-1}$&  &	\footnotesize $2.5 \times 10^{-18}$ \\

\hline

\footnotesize [0.9060, 0.9505] &\footnotesize $< 0.6$ &\footnotesize $\geqslant 94$ & \footnotesize 1.2 $\pm$ 0.8 & \footnotesize 1.3 $\times 10^{-1}$&\footnotesize 0 &	\footnotesize $1.8 \times 10^{-18}$ \\

\hline

\footnotesize [0.9505, 0.9950] &\footnotesize $< 0.6$ &\footnotesize $\geqslant 95$ & \footnotesize 1.2 $\pm$ 0.7 & \footnotesize 1.3 $\times 10^{-1}$&\footnotesize 0 &	\footnotesize $1.5 \times 10^{-18}$ \\

\hline
\end{tabular}
\end{center}
\caption{Results after unblinding of the data (1012 active days live time corresponding to 5 years of data taking). The selection cuts, the number of expected (muons and neutrinos) background and observed events and the upper limits on the flux are presented for each range of velocity ($\beta$). The table was divided into two parts to distinguish the first five bins where $\beta_{fit}$ was assumed as a free parameter from the four bins where $\beta_{fit}=1$.}
\end{table}}

In the first five bins, the reconstructed velocity $\beta_{fit}$ was restricted to be compatible with the range of the MM velocity. Therefore, the event samples in these ranges are exclusive and must be added. 
As shown in Table~1, the sum of background events in the first five ranges adds up to 5.4 events whereas only one event has been observed. This indicates a rather conservative method of extrapolating the atmospheric muon sample into the region defined by the final cuts.
For the last four bins, $\beta_{fit}=1$ and cuts on $\alpha$ and $N_{hit}$ are tightened from bin to bin, that means bin 7 is a subset of bin 6 and so on. Thus, the total background is given here by bin 6 already.

In Fig.~\ref{ul} the ANTARES upper limits as a function of $\beta$ are presented, together with other experimental results from IceCube~\cite{anna}, MACRO~\cite{ambrosio} and Baikal~\cite{aynutdinov}, as well as the previous result from ANTARES~\cite{adrian} and the theoretical Parker bound~\cite{parker}. The MACRO experiment was sensitive also to down-going candidates, surviving the $\sim$3000 meters of water equivalent of the Gran Sasso mountain overburden. Thus, their limit holds for MMs of lower mass (starting from $10^{6}$~GeV/$c^{2}$). For MMs that have to cross the Earth, as in the case of the present paper, the limit is valid for $M > 10^{10}$~GeV/$c^{2}$.
\begin{figure}[!htbp]
\hspace*{-0.6cm}
\centering
\includegraphics[scale=0.7]{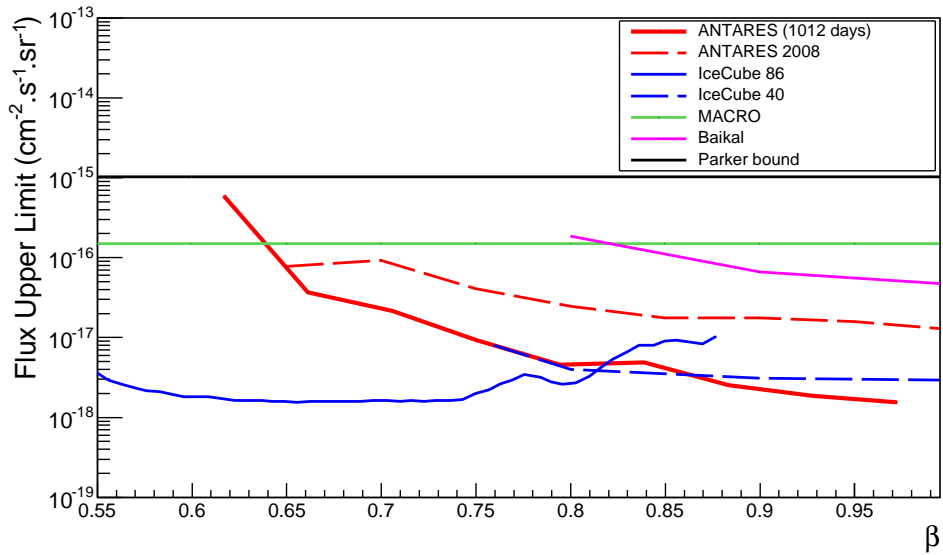} 
\caption{ANTARES 90\% C.L. upper limit on flux for MMs using five years of data with 1012 active days live time (solid red line), compared to the upper limits obtained by other experiments~\cite{anna,ambrosio,aynutdinov}, as well as the previous analysis of ANTARES (dashed red line)~\cite{adrian} and the theoretical Parker bound~\cite{parker}. In~\cite{anna} a more optimistic model for $\delta$-rays production of MMs is used, making a direct comparison difficult.
\label{ul}}
\end{figure}
After applying the final cuts to the unblinded data, two events have been observed. There is one event with $N_{hit}$ = 93, $\alpha$ = 0.5 and zenith = 27.4$^\circ$ which passes the cuts optimized of two bins of $\beta$. It is identified as a bright well-reconstructed neutrino event regarding its physical properties, compatible with the total background observed at this range of high velocities.
The second event with $\beta \geq 0.728$ is consistent with a down-going (zenith = 108.1$^\circ$) atmospheric muon yielding a bright shower.

\section{Conclusion}
 A search for relativistic MMs with the ANTARES neutrino telescope has been performed, using data collected during five years (from 2008 to 2012) and corresponding to a total live time of 1012 days. No signal has been observed above the atmospheric background expectation and new upper limits on the MM flux have been set.

Above the threshold for direct Cherenkov radiation $\beta \geq 0.74$, the limits found are better than those of other neutrino experiments. Below Cherenkov threshold, direct comparison is not straightforward due to the model of cross section used. 

Neutrino telescopes are well suited for the search for MMs. The future detector KM3NeT \cite{km3net} will improve the sensitivity to the detection of MMs due to its large volume and high detection performance.

\setcounter{secnumdepth}{0}
\section{Acknowledgments}

The authors acknowledge the financial support of the funding agencies:
Centre National de la Recherche Scientifique (CNRS), Commissariat \`a
l'\'ener\-gie atomique et aux \'energies alternatives (CEA),
Commission Europ\'eenne (FEDER fund and Marie Curie Program),
Institut Universitaire de France (IUF), IdEx program and UnivEarthS
Labex program at Sorbonne Paris Cit\'e (ANR-10-LABX-0023 and
ANR-11-IDEX-0005-02), Labex OCEVU (ANR-11-LABX-0060) and the
A*MIDEX project (ANR-11-IDEX-0001-02),
R\'egion \^Ile-de-France (DIM-ACAV), R\'egion
Alsace (contrat CPER), R\'egion Provence-Alpes-C\^ote d'Azur,
D\'e\-par\-tement du Var and Ville de La
Seyne-sur-Mer, France;
Bundesministerium f\"ur Bildung und Forschung
(BMBF), Germany; 
Istituto Nazionale di Fisica Nucleare (INFN), Italy;
Stichting voor Fundamenteel Onderzoek der Materie (FOM), Nederlandse
organisatie voor Wetenschappelijk Onderzoek (NWO), the Netherlands;
Council of the President of the Russian Federation for young
scientists and leading scientific schools supporting grants, Russia;
National Authority for Scientific Research (ANCS), Romania;
Mi\-nis\-te\-rio de Econom\'{\i}a y Competitividad (MINECO):
Plan Estatal de Investigaci\'{o}n (refs. FPA2015-65150-C3-1-P, -2-P and -3-P, (MINECO/FEDER)), Severo Ochoa Centre of Excellence and MultiDark Consolider (MINECO), and Prometeo and Grisol\'{i}a programs (Generalitat
Valenciana), Spain; 
Ministry of Higher Education, Scientific Research and Professional Training, Morocco.
We also acknowledge the technical support of Ifremer, AIM and Foselev Marine
for the sea operation and the CC-IN2P3 for the computing facilities.



\end{document}